2012

# The RHIC SPIN Program

## Achievements and Future Opportunities

**Writers, for the RHIC Spin Collaboration**[1]

Elke-Caroline Aschenauer, Alexander Bazilevsky, Kieran Boyle, Kjeld Oleg Eyser, Renee Fatemi, Carl Gagliardi, Matthias Grosse-Perdekamp, John Lajoie, Zhongbo Kang, Yuri Kovchegov, John Koster, Itaru Nakagawa, Rodolfo Sassot, Ralf Seidl, Ernst Sichtermann, Marco Stratmann, Werner Vogelsang, Anselm Vossen, Scott W. Wissink, and Feng Yuan

## Abstract

This document summarizes recent achievements of the RHIC spin program and their impact on our understanding of the nucleon's spin structure, i.e. the individual parton (quarks and gluons) contributions to the helicity structure of the nucleon and to understand the origin of the transverse spin phenomena. Open questions are identified and a suite of future measurements with polarized beams at RHIC to address them is laid out. Machine and detector requirements and upgrades are briefly discussed.

## The Helicity Structure of the Proton

Helicity parton density functions (PDFs) carry vital information on the extent to which quarks and gluons with a given momentum fraction $x$ have their spins aligned with the spin direction of a fast moving nucleon in a helicity eigenstate. The corresponding integrals over all $x$ relate to one of the most fundamental, but not yet satisfactorily answered questions in hadronic physics: how is the spin of the proton distributed among its constituents? The most precise knowledge about helicity PDFs, along with estimates of their uncertainties, is gathered from comprehensive global QCD analyses of available spin-dependent data. Such fits show that while the quark spin contribution is ~30% of the proton spin, contributions from the gluon spin and the orbital angular momenta of quarks and gluon have remained uncertain.

### The polarized gluon distribution $\Delta g$

Longitudinally polarized proton-proton collisions are currently the best source of information on the elusive gluon helicity PDF, $\Delta g$, due to the dominance of gluon induced hard scattering processes. At RHIC, PHENIX and STAR have measured the double helicity asymmetry, $A_{LL}$, of neutral pions and jets, respectively, at mid-rapidity for $\sqrt{s}$ = 200 GeV in 2005 and 2006 [1,2]. These data were included in the DSSV global analysis [3] and significantly constrained $\Delta g(x)$ in the region $0.05 < x < 0.2$. The result for the corresponding integral is $\int_{0.05}^{0.2} \Delta g(x) dx = 0.005 \pm^{0.129}_{0.164}$ at a scale of $Q^2$=10 GeV$^2$, consistent with zero within sizable uncertainties. The uncertainties were determined by mapping out the $\chi^2$ profile for the truncated integral with the robust Lagrange multiplier method. An increase of $\Delta\chi^2/\chi^2 = 2\%$ was identified by DSSV as a conservative estimate of PDF uncertainties [3] and will be used in this document as well.

In 2009, with improved luminosity and polarization, as well as upgraded triggering and data acquisition systems at STAR, both experiments improved considerably the uncertainties in $A_{LL}$ measurements at $\sqrt{s}$ =200 GeV. The impact of these data on global QCD fits based on the DSSV framework is illustrated in Figure 1 and 2. The fit labeled **DSSV+** supplements the RHIC data used in the original **DSSV** analysis with recent results from polarized deep-inelastic scattering (DIS) obtained by COMPASS [4], and the results denoted as **DSSV++** include, in addition, the 2009 RHIC data shown in Figure 1 (left). The preliminary **DSSV++** fit is fully consistent with the previous **DSSV** fit within the uncertainties and shows a preference for a sizable (relative to the total proton spin of $1/2\hbar$) gluon contribution, $\int_{0.05}^{0.2} \Delta g(x) dx = 0.1 \pm^{0.06}_{0.07}$ with significantly reduced uncertainties. Despite this very important achievement, uncertainties for $\Delta g(x)$ remain significant in the presently unmeasured small $x$ region and prevent a reliable determination of the full integral, see Figure 2. By comparing $\Delta g(x)$ integrated in the $x$-range presently covered by RHIC with the integral for the sum of the polarized quark densities $\int_{0.001}^{1} \Delta\Sigma(x) dx = 0.366 \pm^{0.042}_{0.062}$, as determined from DIS data [3], it can be seen that despite the different $x$-ranges gluons can make a very significant contribution to the spin of the proton (Note: $\frac{1}{2} = \frac{1}{2}\Delta\Sigma + \Delta G + L_q + L_g$, with $L_q$ and $L_g$ being the orbital angular momenta of quarks and gluons). To further improve the $\Delta g(x)$ constraint and its integral, we plan to follow three steps: (1) reduce the statistical and systematic uncertainties on the two workhorses of the RHIC $\Delta g$ program, jet and $\pi^0$ $A_{LL}$. (2) Make use of correlation measurements such as di-jets and di-hadrons, which give access to the partonic kinematics and thus the functional form of $\Delta g(x)$ (Note: $x_2 = p_T(e^{\eta_1} + e^{\eta_2})/\sqrt{s}$ and $x_1 = p_T(e^{-\eta_1} + e^{-\eta_2})/\sqrt{s}$, where $\eta_{1,2}$ represent the pseudorapidities of the two outgoing partons). The functional form of $\Delta g(x)$ also provides insight in the dynamical origin of gluons inside the proton. First results from di-jets and di-$\pi^0$ from STAR and PHENIX have been released, and projections for STAR di-jets are shown in Figure 3. (3) Access lower $x$ by performing measurements at $\sqrt{s}$ =500 GeV and at large forward rapidity. Projections for $\pi^0$ $A_{LL}$ measurements in the forward PHENIX calorimeter are shown in Figure 3. With this set of results, we will considerably improve our knowledge of the gluon spin contribution to the proton spin in the *x*-range of 0.002-0.2.

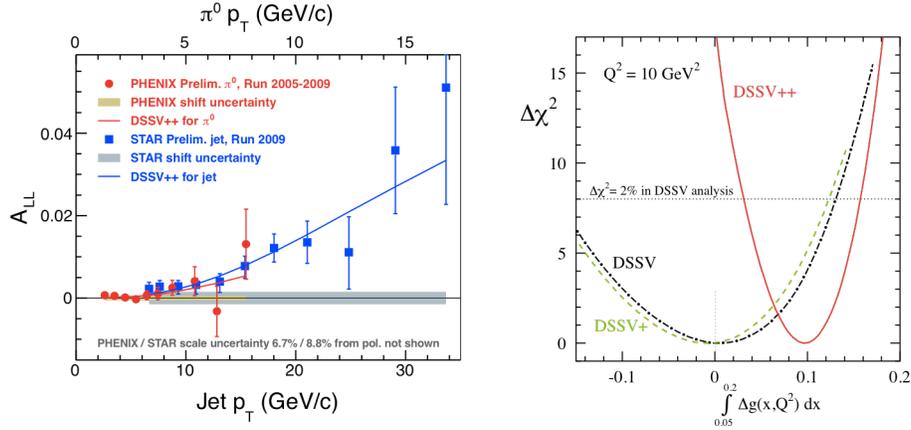

**Figure 1:** Preliminary 2009 data compared to the **DSSV++** fit **(left)** and the $\chi^2$ profile for the integrated gluon contribution in the *x* region currently probed at RHIC for $\sqrt{s}$ = 200 GeV **(right)**. The different $p_T$-scales for $\pi^0$s and jets reflect that an individual $\pi^0$ carries only a fraction of the scattered parton momentum.

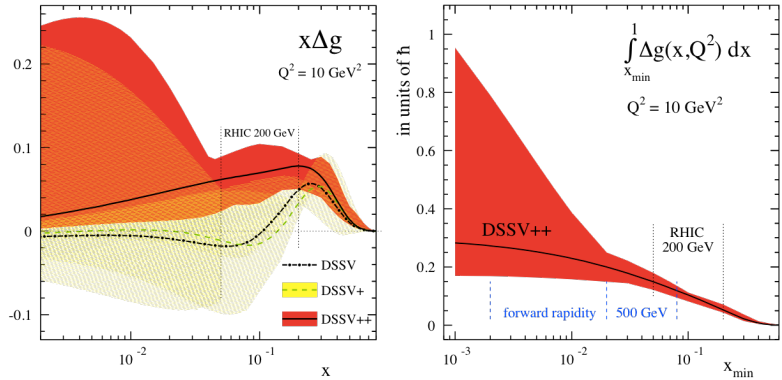

**Figure 2:** Uncertainties in $\Delta g(x)$ with (red band) and without (yellow band) RHIC 2009 data **(left)** and in the integral computed in the range from $x_{min}$ to 1 **(right)** at a scale of 10 GeV$^2$. The flexible functional form for $\Delta g(x)$ used in the DSSV analysis [3] was utilized in all fits.

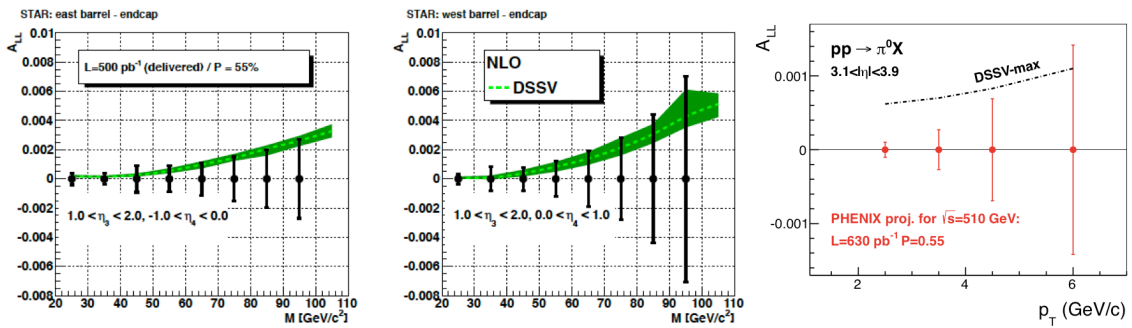

**Figure 3:** Projected statistical uncertainties for $A_{LL}$ of di-jets at mid-forward rapidity at $\sqrt{s}$ = 500 GeV as function of di-jet mass (left, middle) and for electromagnetic clusters (dominated by $\pi^0$s (~80%)) in the PHENIX MPC for $3.1 < \eta < 3.9$ (right). The "DSSV-max" curve corresponds to the upper uncertainty band of **DSSV+** fit in **Figure 2.** Note: the projections quote delivered luminosities.

## The polarized sea quark distributions

The observation of parity violation, even though very well established and tested, is a qualitatively new measurement in polarized $p+p$ scattering at RHIC. The production of $W^{+(-)}$ bosons at $\sqrt{s}$ =510 GeV provides an ideal tool to study the spin-flavor structure of sea quarks inside the proton. The left-handed W boson only couples to (anti)quarks of a certain helicity, giving rise to large parity-violating single spin asymmetries $A_L$ in longitudinally polarized $p+p$ collisions at RHIC. In addition, the coupling of the W's to the weak charge correlates directly to quark flavor. Ignoring quark mixing, $W^{+(-)}$ bosons are produced through $u+\bar{d}$ ($d+\bar{u}$) interactions and can be detected through their leptonic decays $\vec{p} + p \to W^{\pm} + X \to l^{\pm} + X$. Measurements of the transverse momentum and pseudorapidity distribution of the decay leptons can be compared with a recent theoretical calculation at next-to-leading order accuracy [5]. By selecting certain kinematics, the size of the expected asymmetry and the sensitivity to the quark or antiquark helicity, can be enhanced.

The ultimate goal of the W-measurements is to probe the quark and anti-quark spin contribution to the proton spin measured in (SI)DIS at a much higher scale set by the large W mass, $Q^2 \sim 6400$ GeV$^2$ at a medium momentum fraction, $0.05 < x < 0.4$. Do anti-quarks play a decisive role in this? Do $\bar{u}$ and $\bar{d}$ carry similar polarization? These questions become all the more interesting in view of the large difference between the spin-averaged $\bar{u}$ and $\bar{d}$ found in Drell-Yan measurements and predicted earlier [6]. Models of nucleon structure generally make clear predictions about the flavor asymmetry in the sea, $\Delta\bar{u} - \Delta\bar{d} \geq 0$ [7]. This flavor asymmetry in the nucleon is expected to be even larger in the polarized than in the spin-averaged case.

Polarized semi-inclusive DIS measurements are sensitive to the quark and antiquark spin contributions separated by flavor [3]. Dedicated measurements of the quark and anti-quark polarizations have been performed in polarized semi-inclusive DIS by identifying hadrons in the final state. Data have been obtained by the SMC, HERMES, and COMPASS collaborations at scales $Q^2$ ranging from 1 to 50 GeV$^2$ [4,8]. These data are included in the global analysis, labeled **DSSV+** shown in Figure 1 (right) and Figure 5. Indeed the results give first measurements of the polarized light sea quark distributions. This analysis relies on a quantitative understanding of the fragmentation of quarks and antiquarks into observable final-state hadrons, an assumption that is not relevant in the analysis of the $W^{\pm}$ data in polarized proton-proton collisions at RHIC. While the sum of the contributions from quark and antiquark distribution functions (PDFs) of a particular flavor is well constrained in this fit, the uncertainties in the polarized antiquark PDFs separated by flavor remain relatively large.

The STAR preliminary results on $A_L^{W^{\pm}}$ taken during 2012 shown in Figure 4 (left) have been included in the DSSV++ pQCD-fit. Figure 5 (top) shows the results for the $\chi^2$ profile for the truncated integral for $\Delta\bar{u}$ and $\Delta\bar{d}$ in the range $0.05 < x < 1$ at $Q^2$=10 GeV$^2$. A clear improvement on the determination of the polarization of the light sea quarks is observed. For $\Delta\bar{u}$ a shift away from the current best mean value is observed, reflecting that the new STAR $A_L^{W^-}$ data lie above the central curve based on **DSSV+**. Already, with only the preliminary 2012 STAR data, the new global analysis shows a preference for $\Delta\bar{u} > \Delta\bar{d}$ in the range $x > 0.05$. The same fit has been repeated based on the expected uncertainties for $A_L^{W^{\pm}}$ after the 2013-run for both STAR and PHENIX, as shown in Figure 4 (right). A clear further improvement in the truncated integral is observed (solid blue line). The lower panels in Figure 5 show for this fit also the improvement on the polarized light sea quark distributions as functions of $x$.

In summary these results demonstrate that the RHIC W program will in the near-term lead to a substantial improvement in the understanding of the light sea quark polarizations in the nucleon with the expected statistics of the next run.

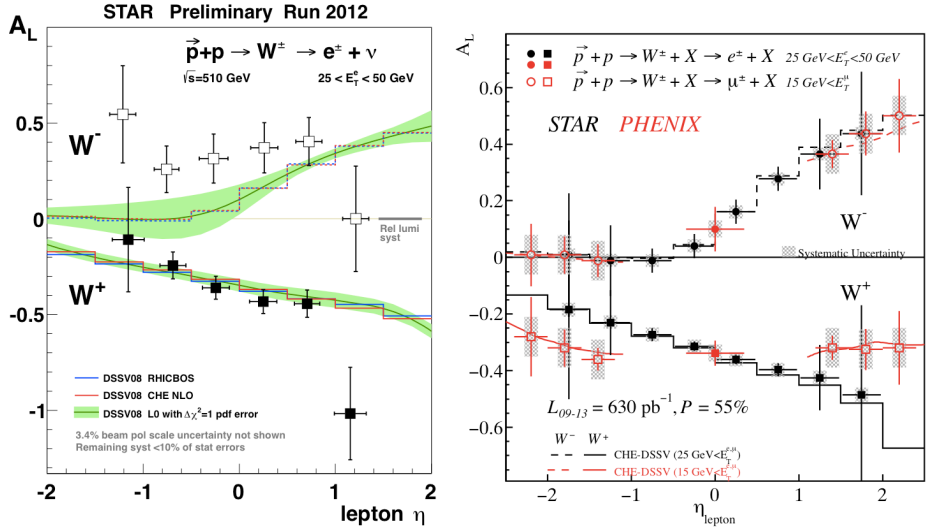

**Figure 4:** $A_L$ for $W^\pm$ as measured by STAR in 2012 **(left)**. Expected uncertainties for $A_L$ for $W^\pm$ for PHENIX and STAR after the 2013 run. The asymmetries have been randomized around the central value of **DSSV** to obtain the results from the pQCD-fit shown in Fig 5 **(right)**. Note: the projections quote delivered luminosities.

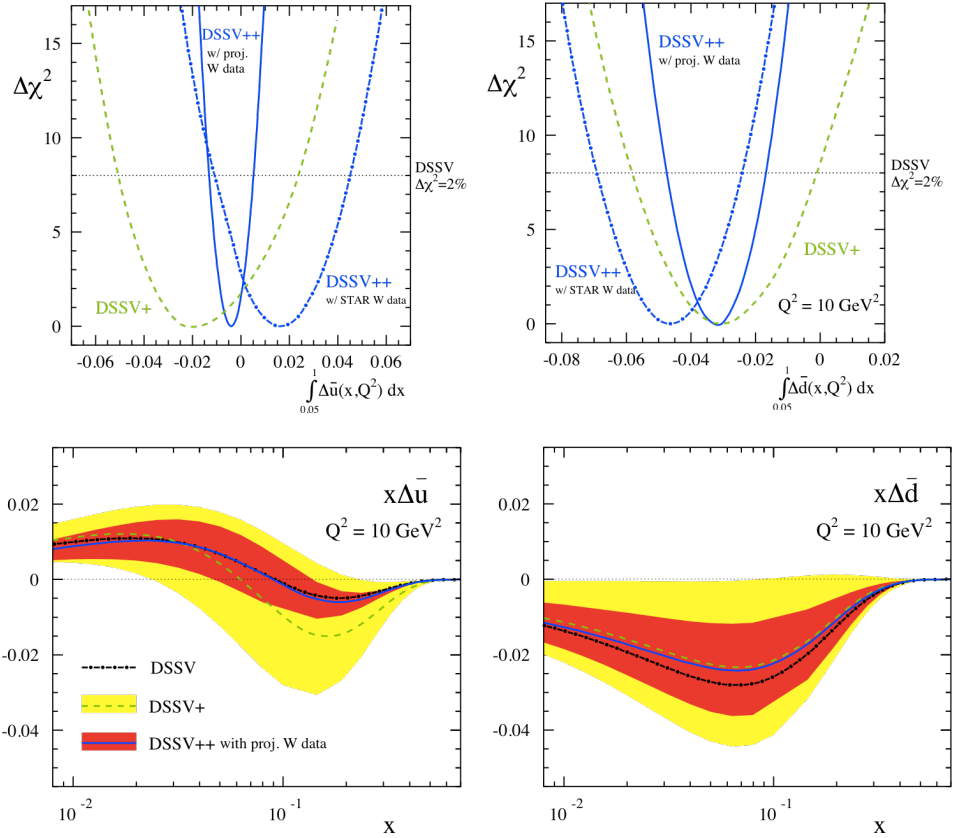

**Figure 5:** $\chi^2$ profiles and $x$-dependent uncertainty estimates for $\Delta\bar{u}$ **(left)** and $\Delta\bar{d}$ **(right)** with (**DSSV++**) and without (**DSSV**, **DSSV+**) including the projections for W boson $A_L$ data shown in the right frame of Fig. 4. In the $\chi^2$ profiles we also illustrate the impact of the preliminary 2012 STAR data from the left frame of Fig. 4 (blue dashed-dotted curves).

## Transverse spin structure of the proton

A natural next step in the investigation of nucleon structure is an expansion of our current picture of the nucleon by imaging the proton in both momentum and impact parameter space. At the same time we need to further our understanding of color interactions and how they manifest in different processes. In the new theoretical framework of transverse momentum dependent parton distributions (TMDs) we can obtain an image in the transverse as well as longitudinal momentum space (2+1 dimensions). This has attracted renewed interest, both experimentally and theoretically in transverse single spin asymmetries (SSA) in hadronic processes at high energies, which have a more than 30 years history. First measurements at RHIC have extended the observations from the fixed-target energy range to the collider regime. Future PHENIX and STAR measurements at RHIC with transversely polarized beams will provide unique opportunities to study the transverse spin asymmetries in Drell-Yan lepton pair, direct photon, and W boson productions, and other complementary processes. Also evolution and universality properties of these functions can be studied. Polarized nucleon-nucleus collisions may provide further information about the origin of SSA in the forward direction and the saturation phenomena in large nuclei at small $x$.

### Transverse asymmetries at RHIC

Single spin asymmetries in inclusive hadron production in proton-proton collisions have been measured at RHIC for the highest center-of-mass energies to date, $\sqrt{s}$=500 GeV. Figure 6 summarizes the measured asymmetries from different experiments as functions of Feynman-$x$ ($x_F \sim x_1-x_2$) and transverse momentum. Surprisingly large asymmetries are seen that are nearly independent of $\sqrt{s}$ over a very broad range. To understand the observed significant SSAs one has to go beyond the conventional collinear parton picture in the hard processes. Two theoretical formalisms have been proposed to generate sizable SSAs in the QCD framework: transverse momentum dependent parton distributions and fragmentation functions, which provide the full transverse momentum information and the collinear quark-gluon-quark correlation, which provides the average transverse information.

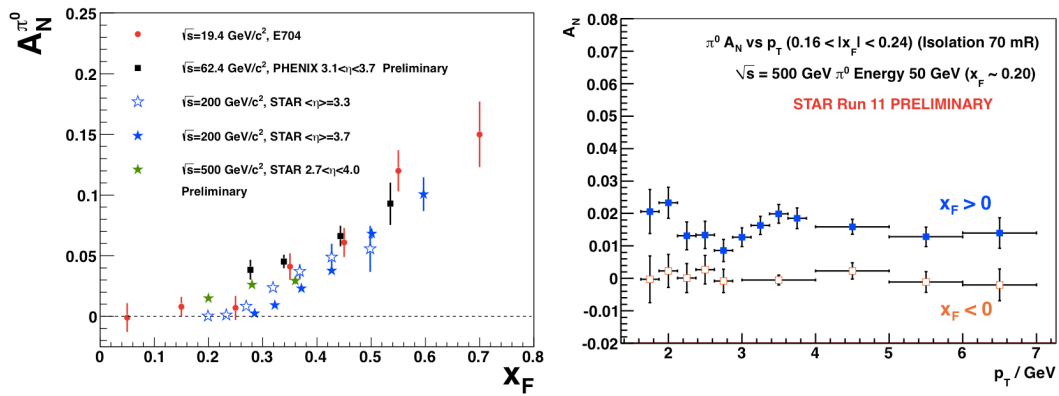

**Figure 6:** Transverse single spin asymmetry measurements for neutral pions at different center-of-mass energies as function of Feynman-$x$ **(left)** and $p_T$-dependence at $\sqrt{s}$ = 500 GeV **(right)**.

At RHIC the $p_T$-scale is sufficiently large to make the collinear quark-gluon-quark correlation formalism the appropriate approach to calculate the spin asymmetries. At the same time, a transverse momentum dependent model has been applied to the SSAs in these hadronic processes as well. Here, various underlying mechanisms can contribute and need to be disentangled to understand the experimental observations in detail, in particular the $p_T$-dependence. These mechanisms are associated with the spin of the initial state nucleon (Sivers/Qiu-Sterman effects) and outgoing hadrons (Collins effects). We identify observables below, which will help to separate the contributions from initial and final states, and will give insight to the transverse spin structure of hadrons.

*Sivers effect*

The Sivers function $f_{1T}^{\perp}$ describes the correlation of parton transverse momentum with the transverse spin of the nucleon. A non-vanishing $f_{1T}^{\perp}$ means that the parton distribution will be azimuthally asymmetric in the transverse momentum space relative to the nucleon spin direction. There is evidence of a quark Sivers effect in semi-inclusive DIS (SIDIS) measurements of the HERMES, COMPASS, and JLab Hall-A experiments [9].

Transverse single spin asymmetries have been measured at RHIC for η, $π^0$ (see Figure 7) and inclusive jets previously [11] at mid-rapidity. Model calculations [12] indicate these asymmetries impose significant constraints on the contributions from the gluon Sivers function (or the three-gluon correlation). Additional data were taken at √s = 200 GeV and 500 GeV during the latest RHIC runs and will allow more precise determination of inclusive jet asymmetries. In addition, the wide $Q^2$-range at RHIC as compared to existing world data from DIS experiments will provide the much-needed input to study the energy and scale dependence for these transverse momentum dependent observables, which is currently under intense theoretical discussion.

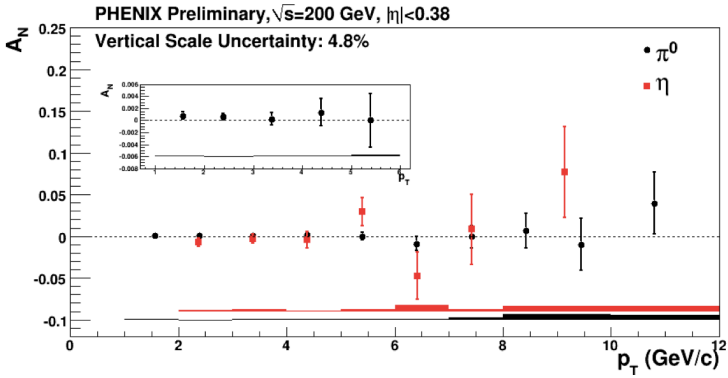

**Figure 7:** Transverse single spin asymmetries of neutral pions and eta-mesons at central pseudorapidity in PHENIX can be used to constrain the gluon Sivers function with input for the respective quark contributions in this kinematic range.

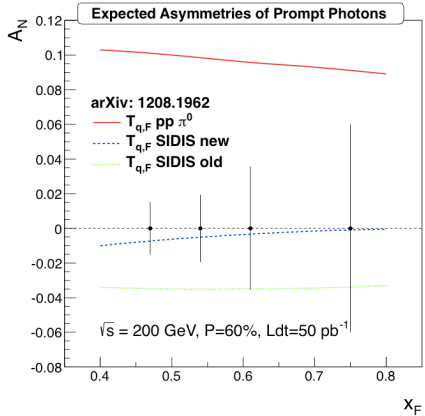

**Figure 8:** Projected errors for prompt photon SSA in the PHENIX MPC-EX. The error bars are combined statistical and systematic errors, including the subtraction of the $π^0$ and η background contributions to the inclusive asymmetry. The blue/green and red curves shown are predictions in the collinear quark-gluon-quark correlation approach based on SIDIS data and *p+p* data, respectively. This measurement will be essential in determining the process contributions to transverse SSA, especially the Sivers function at large *x*. Note: the projections quote delivered luminosities.

The forward direction of the polarized nucleon has played a very important role in our understanding of SSAs. This kinematic region will continue to play an essential role in the future for distinguishing amongst the different underlying mechanisms. Currently STAR has nearly full coverage with electromagnetic calorimeters from -1<η<4. Reconstruction of jets in the regions without charged track information is being investigated. STAR is also investing an upgrade to the inner sectors of the existing Time Projection Chamber (iTPC). This will increase the tracking coverage by about half a unit in rapidity, and therefore allow charged particle and traditional jet reconstruction beyond η>1. PHENIX is in the process of upgrading the forward calorimeters (MPC-EX upgrade) with a preshower detector that will allow the identification of direct photons (see Figure 8) and neutral pions up to energies of 80 GeV at 3.0<$\eta$<3.8.

An important aspect of the Sivers effect, which has emerged from theory lately, is its process dependence and the color gauge invariance. In SIDIS, the quark Sivers function is manifested in association with a final state effect from the exchange of (any number of) gluons between the struck quark and the remnants of the target nucleon. On the other hand, for the virtual photon production in the Drell-Yan process, the Sivers asymmetry appears as an initial state interaction effect. As a consequence, the quark Sivers functions are of opposite sign in these two processes and this non-universality is a fundamental prediction from the gauge invariance of QCD. The experimental test of this sign change is one of the open questions in hadronic physics and will provide a direct verification of QCD factorization.

While the required luminosities for a meaningful measurement of asymmetries in Drell-Yan production are challenging, see Figure 9 (right), other channels can be exploited in $p+p$ collisions, which are similarly sensitive to the predicted sign change. These include prompt photons, $W^{+/-}$ and Z bosons, and inclusive jets and they are already accessible with the existing detectors or modest upgrades and continued polarized beam operations.

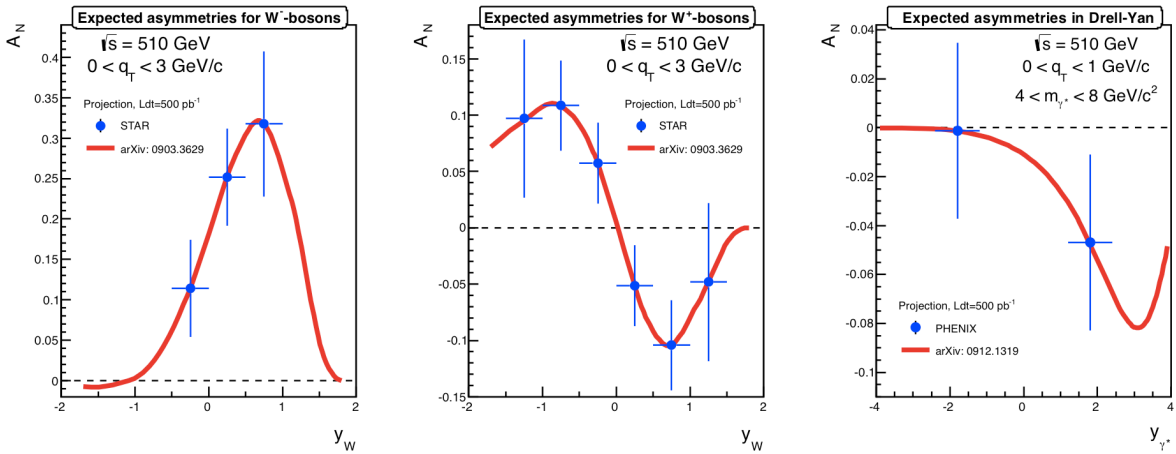

**Figure 9:** Expected uncertainties for transverse asymmetries for reconstructed $W^{\pm}$ **(left, middle)** and Drell-Yan production **(right)** from STAR and PHENIX compared to theoretical predictions based on the SIDIS measurements. These asymmetries provide an essential test for the fundamental QCD prediction of a sign change of the Sivers function in hadronic collisions with respect to SIDIS. Note: the projections quote delivered luminosities.

Predictions for asymmetries of fully reconstructed W/Z bosons are significant in the RHIC kinematic region with a pronounced rapidity dependence [10]. These measurements at very large $Q^2$ would also provide essential input for the evolution effects of the Sivers function as mentioned above.

### *Transversity and Collins effect*

The Collins function $H_1^{\perp}$ describes a correlation of the transverse spin of a scattered quark and the transverse momenta of the fragmentation products and as such can lead to an asymmetry of the distribution of hadrons in jets. Contrary to the Sivers effect discussed above, the Collins fragmentation function is universal among different processes: SIDIS, $e^+e^-$ annihilation, and $p+p$ collisions. This is of special importance to the $p+p$ case where it is always coupled to the chirally odd quark transversity distribution, which describes the transverse spin preference of quarks in a transversely polarized proton. The integral of transversity over $x$ can be compared to lattice QCD calculations of the tensor charge.

Non-vanishing Collins effects have been observed in SIDIS and $e^+e^-$ annihilation [13] and STAR has recently measured Collins asymmetries of charged pions in jets, which establishes the feasibility of the method. The results so far are based on 2.2 pb$^{-1}$ and are dominated by the systematic uncertainties. The introduction of new analysis techniques should allow an order of magnitude reduction in the final systematic uncertainties, both for the 2006 data and the 22 pb$^{-1}$ taken during RHIC run-2012.

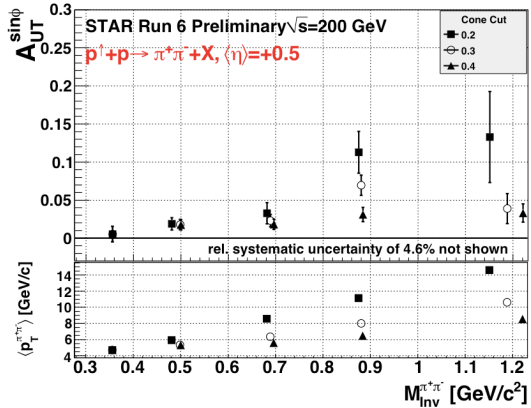

**Figure 10:** Interference fragmentation function asymmetries of di-pions as measured by STAR in 2006 at $\sqrt{s}$ = 200 GeV.

The two-hadron interference fragmentation functions (IFF) have been measured in e+e- annihilation and, in combination with the quark transversity distribution, in deep-inelastic scattering [14]. IFF describes the production of unpolarized hadron pairs in a jet from a transversely polarized quark, and is an interference effect between the hadron pair created in partial waves with a relative angular momentum difference of one. This provides a model independent way to access the quark transversity distribution when the existing data will be compared with similar measurements from RHIC. The preliminary RHIC results show that non-zero asymmetries arise at forward rapidities of the hadron pair especially at increased transverse momenta. The Collins and IFF analyses are being pursued at both $\sqrt{s}$ = 200 GeV (2012) and 500 GeV (2011) leading to a reduction of the statistical errors in Figure 10 by a factor of three. With moderate upgrades, i.e. iTPC, the phase space can be extended into regions that are currently not reached in SIDIS, in particular the high-$x$ region for transversity. This will provide a unique test of the universality of the quark transversity distribution and valuable input for the evolution of TMD fragmentation functions utilizing the above-mentioned Collins asymmetries.

## The longer term RHIC SPIN Program

Both collaborations plan in the 2$^{nd}$ half of the decade detector upgrades to considerably expand their capabilities to address the physics questions discussed above. The upgrades are motivated by spin physics goals of extending the gluon polarization measurements towards smaller gluon fractional momentum values through forward correlation measurements and the study of transverse spin phenomena through precision measurements of SSAs in polarized DY, direct photon production, inclusive jets as well as the Collins and IFF asymmetries described above. The cold nuclear matter and heavy-ion physics goals of these upgrades are described elsewhere [15]. Detailed plans and costs for the upgrades driven primarily by the spin physics program are still under development.

Having a polarized He$^3$ beam available in RHIC and tagging the struck quark flavor through the produced hadron species, will allow for a full quark flavor separation of transverse momentum dependent parton distributions (i.e., transversity and Sivers function). Because of the anti-aligned spins of its two protons, polarized He$^3$ is essentially a surrogate for a polarized neutron beam. Having the additional capability to tag the neutron scattering events by measuring the spectator protons in "Roman Pots" will make the measurement extremely clean. Details on the physics capabilities and all aspects on how to realize a polarized He$^3$ beam in RHIC have been discussed during a workshop in September 2011 and are available online [16].

### *The forward-upgrade to sPHENIX*

The PHENIX Collaboration is in the process of designing a series of upgrades to considerably expand the physics capabilities [17] and make full use of the constantly increasing luminosity at RHIC. The central-rapidity spectrometer will consist of a 2 T solenoid of radius 70 cm surrounded by an electromagnetic calorimeter (EMCal) and a hadronic calorimeter with uniform coverage for $|\eta|<1.1$ to carry out measurements focusing on jets and electromagnetic probes. This is the scope of the currently considered sPHENIX project for DOE. The addition with non-DOE funds of charged particle tracking beyond the existing PHENIX barrel and forward silicon vertex detectors and of a preshower with fine segmentation in front of the electromagnetic calorimeter will allow precision measurements of hadrons (including $\pi^0$s up to $p_T$ = 40 GeV/c) and electrons (di-leptons and from heavy flavor decays).

The open geometry of the magnetic solenoid also allows for the subsequent addition of a forward angle spectrometer aimed at measuring photon, hadron, jet and electron observables. Precision Drell-Yan measurements in the forward spectrometer via di-electrons will require an EMCal, charged particle tracking and heavy flavor tagging for background rejection. The addition of a hadronic calorimeter is highly desirable for jet reconstruction, which will greatly extend PHENIX's ability to measure Sivers, Collins and IFF asymmetries at forward rapidities. Particle tracking will serve to study the dependencies on the fractional momentum of a hadron or a hadron pair in the jet. Particle identification is important for a flavor decomposition of the transverse spin effects.

A "strawman" design of the forward spectrometer in Figure 11 is being used for sensitivity studies and to establish detector requirements to address the physics goals outlined. This spectrometer would cover 1.2 to 4 in pseudorapidity. An extension or modification of the central solenoid is needed to provide a sufficiently strong tracking field in the forward spectrometer, however in the very forward region, $3<\eta<4$, an additional magnetic field would be needed to retain the momentum resolution. The current PHENIX EMCal and MPC are planned to be re-configured for re-use in the upgraded forward spectrometer. It is hoped that the forward upgrade could be available by the end of the current decade.

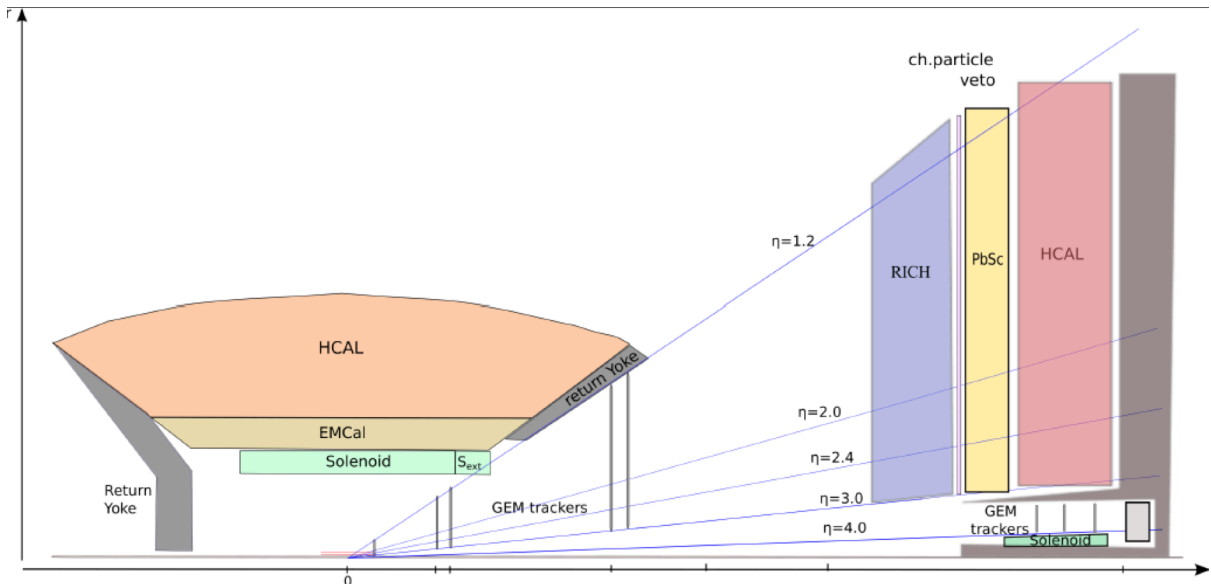

**Figure 11:** Schematic sketch of the sPHENIX central and forward spectrometer layout.

### *The forward STAR upgrade*

The STAR collaboration envisions new instrument upgrades [19] in the second half of the decade to extend the capabilities for measuring jets, electrons, photons, and leading particles, in particular at forward rapidity.

At central rapidities an upgrade is envisioned to the existing Time-Projection Chamber. A key aspect is an increase by a factor two or more of the number of instrumented pad rows for the inner sectors, which will extend the tracking and particle identification capabilities by about half a unit in rapidity. This will serve many areas of the STAR physics program. For spin, this will extend the rapidity coverage with good particle identification, enabling transversity measurements at more forward rapidities which are sensitive to higher-$x$ quarks.

The forward detector upgrade concept includes tracking detectors for charged particles, electromagnetic and hadronic calorimeters, as well as a forward particle identification detector. `Figure 12` shows the schematic design that is being used to quantify detector requirements and measurement sensitivities. A staged approach is envisioned, with the first stage consisting of the Forward Calorimeter System (FCS). The second stage of this upgrade aims at particle separation capability for baryons and mesons. Either threshold Cherenkov or RICH detector technologies are being considered. GEM-based detectors are currently envisioned to provide charged particle tracking.

The FCS is anticipated to be a compensated calorimeter for electron and hadron responses consisting of a front Electromagnetic Calorimeter (EMCal) followed by hadronic calorimeters. The EMCal is expected to adopt a new calorimeter construction technology, using Tungsten-Powder and scintillating fibers. This technology is being developed also as part of an ongoing R&D program aimed at a future Electron-Ion Collider and compact modules with excellent energy resolutions and detector uniformity have been built and tested with beams. The E864 Spaghetti Calorimeter modules will be re-used to provide hadronic calorimetry for part of the forward acceptance. They will be complemented at more forward rapidities with new modules made of lead and scintillator plates. The FCS will thus provide measurement capability for forward jet production and initial Monte Carlo simulations indicate a suitable discrimination of electrons and hadrons, as necessary for example for a Drell-Yan physics program. The aimed for schedule is such that the construction of the FCS will start in the 2$^{nd}$ half of the decade.

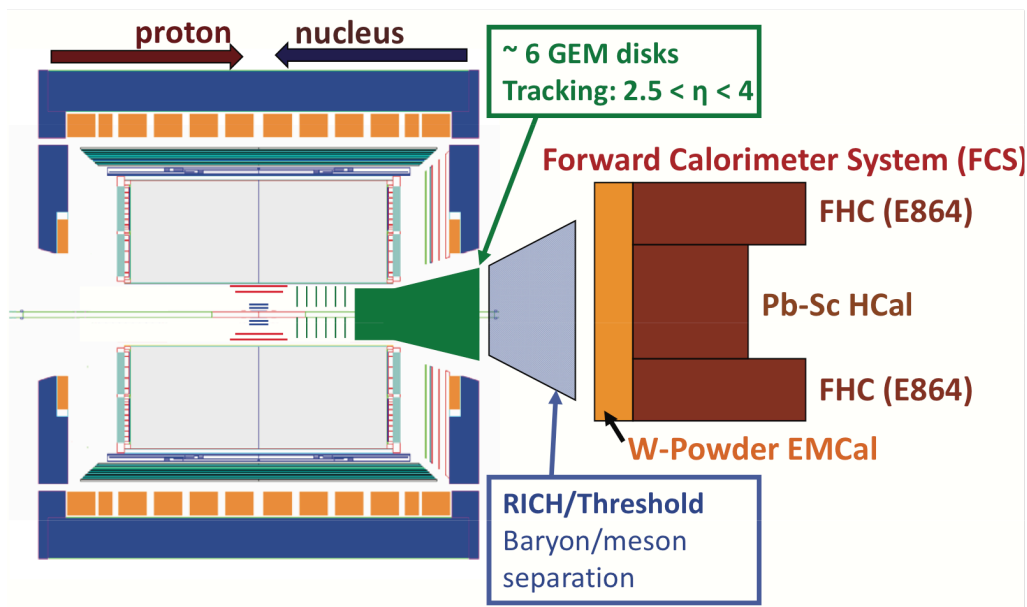

**Figure 12:** Schematic sketch of the STAR forward spectrometer upgrade.

**Polarised Proton-Nucleus Collisions**

*Single Transverse Spin Asymmetry in Polarized Proton--Nucleus Collisions*

As a result of exciting recent theoretical developments, the scattering of a polarized proton on an unpolarized nuclear target appears to have the potential to extend and deepen our understanding of QCD. In the frame where the nucleus is relativistic, its wave function consists of densely packed quarks and gluons, which constantly split and merge with each other. At high enough energies the density of the gluons is so high that the *saturation* regime is reached, characterized by strong gluon fields and scattering cross sections close to the unitarity bound. The saturated wave function is often referred to as the Color Glass Condensate (CGC) and is reviewed in detail in [19-23].

Scattering a polarized probe on this saturated nuclear wave function may provide a unique way of probing the gluon and quark transverse momentum distributions (TMDs). In particular, the single transverse spin asymmetry $A_N$ may provide access to the elusive nuclear Weizsaecker-Williams (WW) gluon distribution function [24,25], which is a solid prediction of the CGC formalism [26-28] but is very difficult to measure experimentally. The nuclear effects on $A_N$ may shed important light on the strong interaction dynamics in nuclear collisions. While the theoretical approaches based on CGC physics predict that hadronic $A_N$ should decrease with increasing size of the nuclear target [29-33], some approaches based on perturbative QCD factorization predict that $A_N$ would stay approximately the same for all nuclear targets [34]. The asymmetry $A_N$ for prompt photons is also important to measure. The contribution to the photon $A_N$ from the Sivers effect [35] is expected to be non-zero, while the contributions of the Collins effect [36] and of the CGC-specific odderon-mediated contributions [33] to the photon $A_N$ are expected to be suppressed [33,37]. Clearly experimental data on polarized proton-nucleus collisions is desperately needed in order to distinguish different mechanisms for generating the single spin asymmetry $A_N$ in nuclear and hadronic collisions.

### *Access to the Generalized Parton Distribution $E_g$*

"How are the quarks and gluons, and their spins distributed in space and momentum inside the nucleon? What is the role of orbital motion of sea quarks and gluons in building the nucleon spin?" These are key questions, which need to be answered to understand overall nucleon properties like the spin structure of the proton. The formalism of generalized parton distributions (GPDs) provide till today the only theoretical framework, which allows some answers to the above questions [38]. The experimentally best way to constrain GPDs is through exclusive reactions in DIS, i.e., deeply virtual Compton scattering. RHIC with its possibility to collide transversely polarized protons with heavy nuclei has world-wide the unique opportunity to measure $A_N$ for exclusive $J/\psi$ in ultra-peripheral $p^\uparrow$+Au collisions (UPC) [39]. A non-zero asymmetry would be the first signature of a non-zero GPD E for gluons, which is sensitive to spin-orbit correlations and is intimately connected with the orbital angular momentum carried by partons in the nucleon and thus with the proton spin puzzle. To measure $A_N$ for exclusive $J/\psi$ in ultra-peripheral $p^\uparrow$+Au collisions provides an advantage in rate as the emission of the virtual photon from the gold nucleus is enhanced by $Z^2$ compared to ultra-peripheral $p^\uparrow$+p collisions. Exclusivity of the process can be ensured by detecting the scattered polarized proton in "Roman Pots" and vetoing the break-up of the gold nucleus.

## Outlook

The physics program described in this document can in its whole only be accomplished over multiple years, as it requires different running conditions:

- transverse or longitudinal beam polarization,
- collision of different beam species $p+p$, $p+\text{He}^3$ or $p+A$,
- different center-of-mass energies.

The table summarizes a possible RHIC run plan taking into account when new machine or detector systems are available that will be needed to perform specific measurements. Only measurements related to the RHIC spin program are listed. It is noted that $p+A$ collisions not only address important questions in spin physics, but also even more importantly give the unique opportunity to study cold QCD matter effects that may serve as backgrounds to quark-gluon plasma phenomena observed in RHIC heavy-ion collisions. The spin program measurements listed in the table are compatible with the more extensive running planned for heavy-ion collisions over the same time periods (see "The Case for Continuing RHIC Operations" [15]).

| Years | Beam Species and Energies | Science Goals | New Systems Commissioned/Required |
|---|---|---|---|
| 2013 | 500 GeV $\vec{p}+\vec{p}$ | Sea antiquark and gluon polarization | Electron lenses upgraded pol'd source |
| 2014 | 200 GeV $p^\uparrow+p$<br><br>200 GeV $\vec{p}+\vec{p}$ | Unravel underlying sub-processes for $A_N$<br><br>Improve precision on $\Delta g(x)$ | PHENIX Muon Piston Calorimeter Extension |
| 2015-2017 | 200 GeV $p^\uparrow+A$<br><br>500 GeV $p^\uparrow+p$<br><br>500 GeV $\vec{p}+\vec{p}$ | Unravel underlying sub-processes for $A_N$, $A_{UT}$ for excl. $J/\Psi \rightarrow$ GPD $E$<br><br>First measurement of PHENIX: $A_N(DY)$, STAR: $A_N(W/Z)$<br>Unravel underlying sub-processes for $A_N$<br>$\Delta g(x)$ at low-$x$, sea antiquark polarizations | STAR inner TPC pad row upgrade |
| >2018 | 200 GeV $p^\uparrow+A$<br><br>160 GeV $p^\uparrow+{}^3He^\uparrow$<br><br>500 GeV $p^\uparrow+p$ | Unravel underlying sub-processes for $A_N$, $A_{UT}$ for excl. $J/\Psi \rightarrow$ GPD $E$<br><br>Quark flavor separation for TMDs<br><br>Precision measurements of transversity, Sivers, IFF, and $A_N(DY)$ | Forward upgrade to sPHENIX STAR forward physics upgrade<br>Polarized He$^3$ beams |

## Summary

A myriad of new techniques and technologies made it possible to inaugurate the Relativistic Heavy Ion Collider (RHIC) at Brookhaven National Laboratory as the world's first high-energy polarized proton collider in December 2001. RHIC delivers polarized proton-proton collisions at center-of-mass energies of up to 500 GeV. This unique environment provides opportunities to study the polarized quark and gluon spin structure of the proton and QCD dynamics at a high energy scale and is therefore complementary to existing semi-inclusive deep inelastic scattering experiments. Recent data from RHIC have for the first time shown a non-zero contribution of the gluons to the proton spin, $\int_{0.05}^{0.2} \Delta g(x)\,dx = 0.1 \pm^{0.06}_{0.07}$ (see Figure 1 and 2).

The results of parity violating single spin asymmetries in W production from 2012 have demonstrated that the RHIC W program will with the expected statistics of the next run lead to a substantial improvement in the understanding of the light sea quark polarizations in the nucleon (see Figure 4 and 5).

In recent years, transverse spin phenomena have gained attention as they offer the unique opportunity to expand our current one-dimensional picture of the nucleon by imaging the proton in both momentum and impact parameter space. At the same time we can further understand the basics of color interactions in QCD and how they manifest themselves in different processes. Results from PHENIX and STAR have also shown that large transverse spin asymmetries for inclusive hadron production that were seen in $p+p$ collisions at fixed-target energies and modest $p_T$ extend to the highest RHIC energies and surprisingly large $p_T$. Future RHIC measurements will provide the essential data to elucidate the dynamical mechanisms that produce these large asymmetries, and crucial tests of the predicted process-dependence of the TMDs.

[1]The **RHIC Spin Collaboration** consists of the spin working groups of the RHIC collaborations, many theorists and members of the BNL accelerator department.